\documentclass[sigconf,nonacm]{acmart}
\usepackage{tikz}

\usepackage{enumitem}
\usepackage[most]{tcolorbox}
\AtBeginDocument{%
  \providecommand\BibTeX{{%
    \normalfont B\kern-0.5em{\scshape i\kern-0.25em b}\kern-0.8em\TeX}}}

\begin{document}

\title{Risk or Chance? Large Language Models and Reproducibility in HCI Research}

\author{Thomas Kosch}
\affiliation{%
  \institution{HU Berlin}
  \city{Berlin}
  \country{Germany}}
\email{thomas.kosch@hu-berlin.de}

\author{Sebastian Feger}
\affiliation{%
  \institution{Rosenheim Technical University of Applied Sciences}
  \city{Rosenheim}
  \country{Germany}}
\email{sebastian.feger@th-rosenheim.de}

\renewcommand{\shortauthors}{Kosch and Feger}

\begin{abstract}

Reproducibility is a major concern across scientific fields. Human-Computer Interaction (HCI), in particular, is subject to diverse reproducibility challenges due to the wide range of research methodologies employed. In this article, we explore how the increasing adoption of Large Language Models (LLMs) across all user experience (UX) design and research activities impacts reproducibility in HCI. In particular, we review upcoming reproducibility challenges through the lenses of analogies from past to future (mis)practices like p-hacking and prompt-hacking, general bias, support in data analysis, documentation and education requirements, and possible pressure on the community. We discuss the risks and chances for each of these lenses with the expectation that a more comprehensive discussion will help shape best practices and contribute to valid and reproducible practices around using LLMs in HCI research.

\end{abstract}



\maketitle

\section{Introduction}
Large Language Models (LLMs) impact and transform most areas of daily life, from education and gaming to creativity and work. Exemplary LLMs are Llama, Alpaca, or GPT-4, with the latter being made accessible to the general public through OpenAI's ChatGPT. Consequently, ChatGPT reached 1 million users within five days after its release and currently has over 180 million users\footnote{\url{https://explodingtopics.com/blog/chatgpt-users}}. In light of the fast-paced developments over the past few years and today's adoption of LLMs across large parts of society, LLMs have also become of great research interest. With the rapid development in natural language processing and today's great accessibility of LLMs to the public, they are anticipated to become a commonplace tool for data analysis in Human-Computer Interaction (HCI) research. For example, LLMs have been explored to accelerate or support the analysis of textual data in HCI~\cite{tabone2023using}, support the user-centered design (UCD) process~\cite{10.1145/3637436}, simulate human samples or replicate user studies~\cite{pmlr-v202-aher23a}, with the latter practices being scrutinized by recent research~\cite{agnew2024illusion}. As with most significant scientific developments, today's high research transparency and validity standards demand a systematic understanding of how the use of LLMs impacts reproducibility. Reproducibility is a major concern across scientific fields. HCI, in particular, is subject to diverse reproducibility challenges due to the wide range of research methodologies employed. The broader scientific community has initiated complex discussions on the potential impact of machine learning and Artificial Intelligence (AI) on reproducibility~\cite{ball2023ai, gibney2022ai}. We aim to contribute to this discourse by focusing specifically on the research and adoption practices of LLMs in the HCI community. This aligns with previous work that advocates for the unique role and responsibility of HCI and human subject research in promoting reproducible practices. Our goal is to design processes and tools that foster scientific reproducibility, thereby enhancing the credibility and validity of HCI research.

In this article, we explore how the increasing adoption of LLMs across all user experience (UX) design and research activities impacts reproducibility in HCI. In particular, we review upcoming reproducibility challenges through the lenses of analogies from past to future (mis)practices, including p- and prompt-hacking, general bias, support in data analysis, documentation, and education requirements, and potential accelerated publication pressure on the community. We discuss the risks and chances for each of these lenses with the expectation that a resulting broader discussion will help shape best practices and contribute to valid and reproducible practices around using LLMs in HCI research.

\section{Large Language Models as Research Tool}
LLMs are built using neural networks, which are computational models inspired by the structure and function of the human brain, using the transformer architecture as a key component. These models are trained on vast amounts of text data, including books, articles, websites, and other written sources. This data teaches the model about language patterns, syntax, semantics, and other linguistic features. The text data is tokenized, broken down into smaller units such as words, subwords, or characters presented numerically, capturing semantic relationships between tokens. In a pre-training phase, the transformer analyzes the tokens at once to decrease a loss function. After pre-training, the model can be fine-tuned on specific tasks or domains to improve performance. Once trained and tuned, the model can be used for various tasks such as text generation, classification, or language translation. Yet, the architecture of LLMs challenges the reproducibility of their outputs. Most LLMs, including GPT-3.5 or GPT-4, are autoregressive models, generating each subsequent token based on preceding tokens within the same sequence. Thus, LLMs cannot holistically self-adjust or validate their outputs. Due to this architectural limitation, LLMs show reduced reasoning and cannot utilize human resources during generation to improve the precision or validation of the output. Although users can use prompt engineering techniques to elicit more thoughtful responses by either integrating specific phrasings within a prompt or conversing over multiple messages with LLMs, these techniques increase the likeliness of unequal outputs when repeating the process. We explain the implications of these limitations in the following.

\subsection{Value Lock-In}
These characteristics complicate the reproducibility of outputs. LLMs are prone to so-called ``value lock-ins,'' meaning that LLMs construct their understanding of human behavior and decision-making by analyzing the norms and attitudes found in human-written texts, such as those found on the internet and in books. However, LLMs usually undergo training only once, missing changes in user standing and opinions that can happen in the future. As a result, the responses they produce may continue to reflect attitudes and beliefs from the time of their initial training. The implications generate research results that appear meaningful, although they may not reflect the facts and beliefs of participants. For example, text-based analyses that comprise interviews or think-aloud data are prone to this. Furthermore, this is impeded by updating the used LLM, where the results can change between versions. While specific parameters, such as the temperature function as magnitude for output ``randomness,'' can be recorded, the exchange of whole models may impact the research results unpredictably. Providing access to previous LLMs, for example, through a repository outlining the LLM version, used parameters, prompts, and outputs, may circumvent some of the described issues.

\subsection{Training Bias}
LLMs are limited to knowledge that is represented in their dataset. ChatGPT, for example, was trained on data available on the internet. Consequently, LLMs are implicitly designed by a fraction of users who share similar properties: persons with access to participate, design, and publish on the internet. LLMs may reflect the views from Western, educated, industrialized, affluent, and democratic backgrounds, reinforcing cultural bias and generating stereotypical representations of marginalized populations when used to analyze research data. Human reviewers can act as secondary observers to investigate the outputs for biases. However, this approach is prone to confirmation biases, where humans seek out, understand, prefer, and remember information that aligns with or reinforces one's existing beliefs or values. Depending on a reviewer's background, this may amplify the bias in LLMs through further confirmation.

\subsection{Hallucination}
Hallucinations have challenged LLMs since their adoption by the public. LLM-based text generation is susceptible to producing unintended content, leading to a decline in the quality of the results. If the results are factually not cross-evaluated with human experimenters, hallucinated results may not be noticed at all and become published. Hallucinations are of great concern when using LLMs to support or simulate human participation processes~\cite{agnew2024illusion}. For example, a lawyer tasked a language model with serving as a legal assistant, resulting in a court filing filled with fictitious legal references. This occurred partly due to the lawyer's trust regarding the authenticity of the referenced cases, which could be revealed through factual testing. How would qualitative and quantitative analysis results be tested in the context of HCI research? The HCI research community must validate methods that test for hallucinations in HCI research results.

\section{Implications for Reproducibility in HCI Research}
HCI research is characterized by a methodological diversity in designing and evaluating systems that pose various reproducibility challenges already today \cite{feger_role, chat_transparency}. The expected widespread adoption of LLMs as part of ideation support \cite{shaer2024aiaugmented}, the substitution of human participants for design requirement mapping and system evaluations \cite{agnew2024illusion, 10.1145/3637436}, and as support for data analysis \cite{tabone2023using} open up an entirely new spectrum of reproducibility challenges in HCI research. We map risks and chances for HCI reproducibility and discuss initial recommendations across these different phases of UX research.

\subsection{Learning from Today's Reproducibility Challenges: Let's not Repeat Mistakes}

Numerous factors contribute to the reproducibility challenges we face today. In quantitative research, one major issue relates to p-hacking. P-hacking refers to the selective reporting of statistical tests to achieve statistically significant results, leading to inflated false positives and compromised reproducibility. This practice not only undermines the integrity of scientific research but also perpetuates a cycle of erroneous findings. While moving towards the increasing adoption of LLMs in HCI research, we have to consider how to address existing reproducibility challenges and carefully navigate new pitfalls that might arise from LLMs. Analogous to p-hacking, using LLMs during UX research is vulnerable to the so-called ``prompt hacking.'' Prompt hacking of LLMs mirrors p-hacking in research by manipulating inputs to influence outputs. As p-hacking selectively reports statistical tests to achieve desired results, prompt hacking skews LLM responses by adjusting input prompts. Both practices compromise integrity: p-hacking distorts scientific findings while prompt hacking biases language model outputs. Recognizing these parallels highlights the importance of transparency and integrity in research and AI development, urging us to prioritize robust methodologies to uphold credibility and reliability. Yet, the community can learn from the past to avoid similar mistakes with new technologies in the future. We propose the following to avoid repeating mistakes:
\begin{enumerate}[wide, labelwidth=0pt, labelindent=0pt]
        \item We argue that adopting LLMs as part of UX research must consider reproducibility challenges and specifically identify analogous issues introduced by LLMs.\\
        \item We suggest learning from current best practices and advocating their adoption for the use of LLMs. Regarding the examples of p-hacking and prompt-hacking, we note that the applicability of established tools like pre-registration and transparent, prompt protocols as part of manuscript submissions and paper publications should be evaluated.
\end{enumerate}

\subsection{Bias Across User Experience Research}

UX research and system evaluation fundamentally require knowledge about human perception and experience. In this regard, HCI research today faces the issue of bias through sampling experiences of too few or a biased set of human samples. In this context, Schmidt et al. \cite{10.1145/3637436} stressed that the ``basic idea is that LLMs encode human experiences, which may be drawn upon in design.'' Expanding on this notion, we perceive a substantial opportunity for LLMs to increase information's robustness and consequent reproducibility throughout the UX research process, from requirements mapping to system evaluation. However, this approach also represents reproducibility risks. Today, HCI research is confronted with the criticism that findings often reflect the perspectives of young, educated, and Western communities. LLMs can further aggravate this problem if they are sampled on reports and experiences from a distinct part of society. Many experiences will reflect technology-savvy individuals with a tendency for a younger population. Further, LLMs are likely to favor specific languages and cultures. Any transparency regarding these training biases further makes assessment of risks difficult. Concerning bias and its impact on research reproducibility, we map multiple requirements:
\begin{enumerate}[wide, labelwidth=0pt, labelindent=0pt]
        \item HCI should support the development of LLMs to make the selection of training data and subsequent biases transparent. \\
        \item The HCI community should use multiple LLMs across UX research to reflect broad human perspectives wherever necessary. Combining LLMs that lean towards different regions or cultures is an example. \\
        \item Research and reviewers must critically examine the interconnections between LLM training transparency, combinations of different LLMs, interplay with additional direct human subject reporting, and the resulting deviation for the specific activity and use case.
\end{enumerate}

\subsection{LLMs for Cross-Validation and Analysis Support}
HCI is subject to a rich diversity of research methods unmatched in most other fields. While we appreciate this richness, it comes with various reproducibility issues subject to specific methods, which generally impact HCI research reproducibility. Many HCI researchers are involved in activities across multiple methodologies. As it is difficult to gain the same expertise across all methods, individual researchers might find it more challenging to evaluate and counteract reproducibility issues for some of the methods they employ. We see an opportunity for LLMs to help educate researchers about reproducibility pitfalls across methods and support the validation of research findings. For example, regarding qualitative methods, which have been subject to claims of reproducibility, LLMs might provide additional verification complementing manual data analysis. This also holds an opportunity to counter the bias of a single or few interpreters analyzing qualitative research data, as Tabone and de Winter \cite{tabone2023using} demonstrated. Similarly, LLMs can support quantitative data analyses by cross-checking and reasoning about the applicability of statistical tests and the validation of calculations through a second entity. At the same time, we note the risk of overreliance on LLMs across HCI research methods. LLMs can help improve research reproducibility as an assistive tool across various HCI research methods. Besides providing cross-validation, LLMs can help fill individual gaps in best practices. Yet, LLMs should carefully be used as a supportive tool rather than a single source of analysis, bearing the risk of overreliance, bias, and subsequently irreproducibility.

\subsection{Defining New Reporting Requirements and Educating the Community}

More documentation of data and metadata must be required to represent a key issue for research reproducibility today. Various initiatives across research fields attempt to encourage or demand scientists to provide the most accurate documentation of recording conditions, hardware setups, software specifications, and many more. The introduction of LLMs into research practices requires developing new best practices regarding the reporting of LLM usage. Diverse fields like HCI must specifically investigate and pose requirements that apply across research methods. We suggest the following:
    \begin{enumerate}[wide, labelwidth=0pt, labelindent=0pt]
        \item Establish precise documentation requirements as part of publication venues that demand detailed information regarding the scope of LLM use, prompts entered, and corresponding metadata like concrete LLMs used and their specific versions.  \\
        \item Provide accessible educational resources for the wider HCI research community that explain those reporting requirements and general LLM use challenges such as bias, hallucination, and value lock-in. \\
        \item Contribute to the development and incentivize the use of transparent and accessible LLMs that provide detailed information on the type of data used for training and that remain accessible to the community for reproduction. This targets the primary reproducibility concern that older or specific versions of commercial and proprietary LLMs become unavailable to the public.
    \end{enumerate}

\subsection{The Risk of Increased Research Pressure on HCI Reproducibility}
Researchers face the challenge of producing high-quality research while ideally generating sufficient output to advance their careers. The use of LLMs provides an opportunity to increase efficiency. For example, Shaer et al. \cite{shaer2024aiaugmented} demonstrated that LLMs can support creative ideation processes. Along those lines, Schmidt et al. \cite{10.1145/3637436} envisioned the partial substitution of human participants through LLMs, and Tabone and de Winter \cite{tabone2023using} discussed using LLMs in data analysis and reporting. These opportunities might create pressure, as Schmidt et al. \cite{10.1145/3637436} underlined: ``By using LLMs, we might make UCD cheaper and hence more widely applicable; at the same time, though, we put pressure on the field to move this way to stay competitive. Hence, the transparency about how UCD is conducted and to what extent models are used is critical.'' Considering the concern of research reproducibility, we must also address the potential risks associated with the pressure to adopt LLMs. The premature introduction of LLMs into general HCI research could manifest mispractices before best practices across the diverse HCI methodologies can be established and communicated. Moreover, the anticipated increase in paper submissions could strain an already busy peer review system, potentially impacting the time reviewers can dedicate to assessing the reproducibility implications of LLM use. We suggest the following:
    \begin{enumerate}[wide, labelwidth=0pt, labelindent=0pt]
        \item Manage and communicate expectations and address concerns regarding LLM use and potentially perceived pressure within and across HCI laboratories.   \\
        \item Develop, communicate, and demand best practices as quickly as possible as part of paper submission requirements. This might happen through dedicated workshops and panels at HCI publication venues. \\
        \item Educate peer reviewers about best practices and evaluate whether and how LLMs might support increasingly complex and rich peer review processes, considering potential pitfalls like bias, hallucination, and value lock-in.
    \end{enumerate}

\section{Risk or Chance for Reproducibility in HCI?}
LLMs present a significant opportunity for HCI research to accelerate data analysis and disseminate results. While acknowledging the benefits of employing LLMs to make HCI data analysis procedures more accessible to the community, it is essential to approach their increasing use cautiously. The structure, training methodologies, and frequent updates of LLMs are not designed to yield consistent results, which can affect reproducibility, manifest biases, and increase pressure on publication processes. This article suggests a discourse on the influence of LLMs within the HCI community and their impact on research reproducibility. We aim to generate interest in a series of focused discussions we plan to organize soon. These conversations will explore establishing specialized scientific platforms for disseminating research conducted in this domain, aiming to maintain reproducibility when employing AI tools for HCI data analysis.

\bibliographystyle{ACM-Reference-Format}
\bibliography{main}

\end{document}